\documentclass[preprint,12pt]{elsarticle}

\usepackage{amssymb}
\usepackage{amsmath}
\usepackage{enumitem}
\usepackage{tabularx}
\usepackage{multirow}
\usepackage{booktabs}
\usepackage{graphicx}
\usepackage{makecell}
\usepackage{xcolor}
\usepackage{colortbl}

\journal{Fundamental Research}
  
\begin{document}
  
\begin{frontmatter}

  \title{Finite-Blocklength Information Theory} 

\author[label1,label2]{Junyuan Gao} 
\affiliation[label1]{organization={Department of Electrical and Electronic Engineering, The Hong Kong Polytechnic University},
            city={Hong Kong SAR},
            country={China}
            }
\affiliation[label2]{organization={Department of Electronic Engineering, Shanghai Jiao Tong University},
            city={Shanghai},
            postcode={Minhang 200240}, 
            country={China}}

\author[label2]{Shuao Chen} 

\author[label2]{Yongpeng Wu\corref{cor1}} 

\cortext[cor1]{Corresponding author}
\ead{yongpeng.wu@sjtu.edu.cn}

\author[label1]{Liang Liu} 

\author[label3]{Giuseppe Caire} 

\affiliation[label3]{organization={Communications and Information Theory Group, Technische Universit{\"a}t Berlin},
            city={Berlin},
            postcode={10587}, 
            country={Germany}} 

\author[label4]{H. Vincent Poor} 

\affiliation[label4]{organization={Department of Electrical and Computer Engineering, Princeton University},
            city={Princeton},
            postcode={NJ 08544}, 
            country={USA}}

\author[label2]{Wenjun Zhang} 

\begin{abstract}
  Traditional asymptotic information-theoretic studies of the fundamental limits of wireless communication systems primarily rely on some ideal assumptions, such as infinite blocklength and vanishing error probability. 
  While these assumptions enable tractable mathematical characterizations, they fail to capture the stringent requirements of some emerging next-generation wireless applications, such as ultra-reliable low latency communication and ultra-massive machine type communication, in which it is required to support a much wider range of features including short-packet communication, extremely low latency, and/or low energy consumption. 
  To better support such applications, it is important to consider finite-blocklength information theory. 
  In this paper, we present a comprehensive review of the advances in this field, followed by a discussion on the open questions. 
  Specifically, we commence with the fundamental limits of source coding in the non-asymptotic regime, with a particular focus on lossless and lossy compression in point-to-point~(P2P) and multiterminal cases. 
  Next, we discuss the fundamental limits of channel coding in P2P channels, multiple access channels, and emerging massive access channels.  
  We further introduce recent advances in joint source and channel coding, highlighting its considerable performance advantage over separate source and channel coding in the non-asymptotic regime. 
  In each part, we review various non-asymptotic achievability bounds, converse bounds, and approximations, as well as key ideas behind them, which are essential for providing engineering insights into the design of future wireless communication systems.


\end{abstract}

\begin{keyword}
  approximation \sep finite-blocklength information theory \sep low latency \sep non-asymptotic bound \sep source and channel coding  
  
  
  
  \end{keyword}

\end{frontmatter}

\section{Introduction}

As one of the three main use cases in fifth-generation (5G) wireless networks, ultra-reliable low latency communications (URLLC) forms the foundation for key applications that require strict end-to-end delay and reliability~\cite{ITU_M2160}. Examples include industrial automation, autonomous driving, remote healthcare, augmented reality (AR) and virtual reality (VR). 
To support these applications, the 3rd generation partnership project (3GPP) has specified typical URLLC requirements of $1$ ms physical layer latency and 99.999\% reliability for 5G networks~\cite{3GPP1,3GPP2}. 
In the upcoming sixth-generation (6G) wireless networks, latency is expected to shrink from milliseconds to sub-milliseconds or even microseconds and reliability is expected to rise from 99.999\% to 99.9999\%~\cite{you2023toward,pourkabirian2024vision}. URLLC will play a key role in supporting a wider range of emerging applications, including both the further development of 5G applications and new scenarios such as real-time human-machine interaction, fully autonomous driving, and human-centered immersive communications~\cite{huang2024challenges,li2024cellular}. More reliable communication at shorter blocklengths is key to achieving URLLC.

Shannon's asymptotic information-theoretic results characterize the fundamental limits of wireless communication systems primarily relying on some ideal assumptions, such as infinite blocklength, infinite payload size, and vanishing error probability~\cite{Shannon}. 
These assumptions enable tractable mathematical characterizations of the minimum achievable coding rate for source coding and the maximum achievable coding rate for channel coding. 
However, these assumptions fail to capture the stringent requirements of many emerging applications mentioned above, in which it is required to support a much wider range of features including short-packet communication, extremely low latency, and small but non-negligible error probability. 
The mismatch between the ideal assumptions and the features of practical systems exposes the limitations of asymptotic information theory in characterizing the fundamental limits of practical communication systems. 
Therefore, it is essential to explore rigorous non-asymptotic frameworks -- a pursuit demanding novel analytical tools and techniques to address the challenges in the finite-blocklength regime.  

\begin{figure}[t]
  \centering
  \includegraphics[width=1\linewidth]{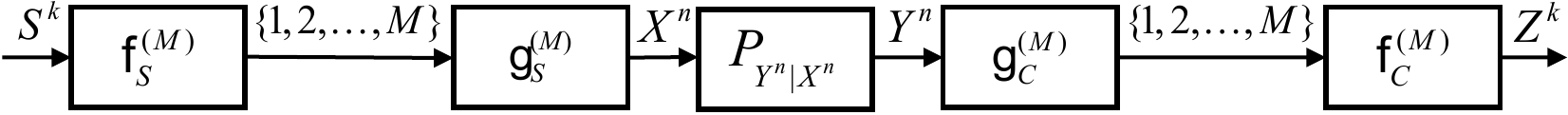}
  \caption{Source coding and channel coding setup.} \label{Fig:source channel coding}
\end{figure}

Finite-blocklength information theory has received significant attention in recent years, covering source coding, channel coding, and joint source-channel coding (JSCC). 
The source-channel coding setup is shown in Fig.~\ref{Fig:source channel coding}. 
For source coding or data compression, the source output, a length-\(k\) sequence, is mapped to a bit sequence from the set \(\{1,\ldots,M\}\) so that at the receiver the source symbols can be recovered exactly (for almost lossless compression) or within a certain distortion fidelity (for lossy compression), where we are given a source alphabet \(\mathcal{S}\), a reconstruction alphabet \(\mathcal{Z}\) and a distortion measure \(\mathsf{d}: \mathcal{S} \times \mathcal{Z} \mapsto [0, +\infty]\).
Finite-blocklength information theory for source coding focuses on the tradeoff among the blocklength \(k\), the error probability—either \(\mathbb{P}[S\neq Z]\) in the (almost) lossless case for a discrete source \(S\)~\cite{kontoyiannis2013optimal}, or \(\mathbb{P}[\mathsf{d}(S,Z)>d]\) in the lossy case—and the coding rate \(\log M/k\)~\cite{kostina2012fixed}. 
The basic task of channel coding is to transmit \(M\) messages with blocklength \(n\) over a noisy channel so that they can be distinguished with error probability below \(\epsilon\) at the receiver~\cite{polyanskiy2010channel}. In this case, the fundamental problem lies in characterizing the tradeoff among blocklength \(n\), error requirement \(\epsilon\), and data rate \(\log M / n\). 
In most contemporary communication systems, the above mentioned source coding and channel coding tasks are performed sequentially, which is known as SSCC. 
However, this architecture suffers from significant performance loss at finite blocklengths, calling for joint design of both source and channel encoders and decoders. 
In such a JSCC scheme, the source produces a \(k\)-length sequence that is directly mapped to an \(n\)-length sequence suited for channel transmission, and thus the data rate is given by \(k/n\). The fundamental tradeoff between \(k\), \(n\), and the error probability is of great significance~\cite{kostina2013lossyJSCC}. 

In the finite-blocklength regime, the exact description of the tradeoffs mentioned above is analytically intractable. As a result, researchers turn to develop tight and computable achievability bounds, converse bounds, and approximations as follows: 
\begin{itemize}[leftmargin=18pt]
  \item Achievability bound:  A triple \((M, k \text{ or } n, \epsilon)\) is said to be achievable if there exists a code of size \( M \), blocklength \( k \) for source coding or \( n \) for channel coding, and error probability \( \epsilon \). More generally, for JSCC, a triple \((k, n, \epsilon)\) is achievable if there exists a joint source-channel code with source input length \( k \), channel blocklength \( n \), and error probability \( \epsilon \). An achievability bound provides an inner bound on the achievable region, establishing the existence of codes for a subset of parameters \((M, k \text{ or } n, \epsilon)\) or \((k, n, \epsilon)\). 
  \item Converse bound: A converse bound provides a non-existence result, showing that no code exists for certain values of the parameters \((M, k \text{ or } n, \epsilon)\) or \((k, n, \epsilon)\). 
  When the converse region coincides with the complement of the achievability region, we have a tight result, i.e., we have determined the largest possible region. 
  \item Approximation: In many cases, a tight characterization of the region is only possible in some asymptotic regime (e.g., typical regime is \( k \text{ or } n \to \infty \) and \( \epsilon \to 0 \)). However, it is often possible to describe the scaling laws of the region in terms of dominant terms (possibly up to constants) for all non-asymptotic values of \( M \), \( k \text{ or } n \), and \( \epsilon \). Additionally, using tools like the law of large numbers and ergodic theorems, one can derive easily computable approximations for the error probability \( \epsilon \) of optimal codes for given \((M, k)\) in souce coding or \((M, n)\) in channel coding or \((k, n)\) in JSCC.  
\end{itemize}

For recent advances in finite-blocklength results of lossy source coding, readers may refer to the monograph by Zhou and Motani~\cite{zhou2023finite}, where the fundamental limits of both point-to-point (P2P) and multiterminal scenarios were discussed. On the channel coding side, the seminal work of Polyanskiy, Poor, and Verdú~\cite{polyanskiy2010channel} established fundamental non-asymptotic bounds that have inspired a large body of follow-up research. The work of Kostina and Verdú~\cite{kostina2013lossy} provided comprehensive finite-blocklength characterizations for lossy source coding and offered valuable insights into the comparison between separate source-channel coding (SSCC) and JSCC. 
More related works will be reviewed in the subsequent sections. 
To provide an organized overview of the main results discussed throughout this paper, we summarize them in Table~\ref{tab:summary}.
These finite-blocklength results characterize the fundamental limits of wireless communication systems with stringent latency constraints, thereby providing benchmarks to evaluate practical wireless communication schemes and identifying the scenarios with potential for further performance improvement.

\begin{table}[!t]
\centering
\caption{Summary of key results in the finite-blocklength regime}
\vspace{2.5mm}
\resizebox{\textwidth}{!}{%
\renewcommand{\arraystretch}{1.8} 
\begin{tabular}{|c|c|c|c|}
\hline
\textbf{Category} & \multicolumn{2}{c|}{\textbf{Subcategory / Problem}} & \textbf{Reference} \\
\hline
\multirow{6}{*}{\makecell{Source \\ Coding}} 
& \multicolumn{2}{c|}{P2P Lossless Compression} & \cite{kontoyiannis2013optimal,verdu2007teaching} \\ \cline{2-4}

& \multirow{3}{*}{\makecell{P2P Lossy \\ Compression}} 
& General Memoryless Source & \cite{kostina2012fixed,kontoyiannis2000pointwise,marton1974error,han2006reliability,iriyama2001probability,yang1999redundancy} \\ \cline{3-4}
& & Gauss-Markov Source & \cite{gray1970information,tian2019dispersion,tian2021nonstationary} \\ \cline{3-4}
& & Mismatch & \cite{lapidoth1997role,zhou2018refined} \\ \cline{2-4}

& \multirow{2}{*}{\makecell{Multi-terminal \\ Coding}} & Kaspi Problem & \cite{kaspi2002rate,zhou2019non} \\ \cline{3-4}
& & Successive Refinement & \cite{no2016strong,zhou2017second,rimoldi1994successive,effros1999distortion} \\ \hline

\multirow{3}{*}{\makecell{Channel \\ Coding}} 
& \multicolumn{2}{c|}{P2P} & \cite{polyanskiy2010channel,yang2018beta,polyanskiy2011feedback,yavas2025variable,Yury_conf1,Yang_quasi,Yury_ISIT,Collins,Hoydis,You_TWC,Yury_conf2,Altug}  \\ \cline{2-4}
& \multicolumn{2}{c|}{Multiple Access} & \cite{MolavianJazi,huang2012finite,haim2012note,Scarlett,scarlett2015second} \\ \cline{2-4}
& \multicolumn{2}{c|}{Massive Access} & \cite{A_perspective_on,RAC_fading,han2025finite,noKa,myTIT,letter,Gao_TIT2} \\ \hline

\multirow{2}{*}{JSCC} 
& \multicolumn{2}{c|}{Theory} & \cite{kostina2013lossyJSCC,csiszar1982error,pilc1968transmission,wyner1972transmission,csiszar1981graph,wang2011dispersion} \\ \cline{2-4}
& \multicolumn{2}{c|}{Applications} & \cite{gunduz2024joint,bourtsoulatze2019deep,kurka2020deepjscc} \\ \hline

\end{tabular}%
}
\label{tab:summary}
\end{table}

The paper is organized as follows. 
In Section~\ref{sec:sc}, we review the finite-blocklength limits of source coding. In Subsection~\ref{subsec:Preliminaries}, we provide definitions of some essential terms to ensure clarity and consistency in the subsequent discussion.
Specifically, in Sections~\ref{subsec:lossless sc}, \ref{subsec:lossy sc}, and \ref{subsec:multiterminal sc}, we cover lossless and lossy compression in point-to-point and multiterminal settings. In Section~\ref{sec:cc}, we review the finite-blocklength limits of channel coding with discussions on point-to-point channels, multiple access channels, and emerging massive access channels in Sections~\ref{subsec:pp cc}, \ref{subsec:mac cc}, and \ref{subsec:mra cc}, respectively. In Section~\ref{sec:jscc}, we review the finite-blocklength limits of JSCC. Open problems and future research directions in finite-blocklength information theory are presented in Section~\ref{sec:open} and we conclude this paper in Section~\ref{sec:conclusion}.

\section{Source Coding}
\label{sec:sc}

In information theory, source coding, also known as data compression~\cite{wade1994signal}, is broadly divided into two types: \textit{almost lossless compression} and \textit{lossy compression}. Almost lossless compression is typically applied to discrete sources and ensures that all original information is preserved, enabling perfect reconstruction. Lossy compression applies to both discrete and continuous sources. It achieves higher efficiency by discarding non-essential information, which approximates the original data within a specified distortion level.  
In this section, we review the advances in the fundamental limits of lossless compression and lossy compression in P2P settings, respectively. Then, the fundamental limits in multiterminal settings are discussed. 

\subsection{Preliminaries}
\label{subsec:Preliminaries}
To ensure clarity and standardization, we introduce some commonly used terminologies and definitions here.
In fixed-length lossy compression, a general source with alphabet \( \mathcal{S} \) and distribution \( P_S \) is mapped to one of \( M \) codewords from a reproduction alphabet \( \mathcal{Z} \). A lossy code consists of two possibly randomized mappings: \( \mathsf{f}: \mathcal{S} \mapsto \{1, \dots, M\} \) and \( \mathsf{g}: \{1, \dots, M\} \mapsto \mathcal{Z} \). The performance of such a code is evaluated using a distortion measure \( \mathsf{d}: \mathcal{S} \times \mathcal{Z} \mapsto [0, +\infty] \). Given a decoder \( \mathsf{g} \), the optimal encoder assigns each source output \( s \) to the codeword minimizing the distortion, i.e., \( \mathsf{f}(s) = \arg\min_{m} \mathsf{d}(s, \mathsf{g}(m)) \). The average distortion over the source statistics is commonly used as a performance metric. Additionally, the probability of exceeding a specified distortion level, termed the excess-distortion probability, provides a stricter criterion for evaluation.  

An \( (M, d, \epsilon) \) code for \( \{\mathcal{S}, \mathcal{Z}, P_S, d\} \) is a code with \(|\mathsf{f}|=M\) such that \( \mathbb{P}[\mathsf{d}(S, \mathsf{g}(\mathsf{f}(S))) > d] \leq \epsilon \). The smallest code size \( M^\star \) achievable at distortion \( d \) and excess-distortion probability \( \epsilon \) is defined as 
\begin{align}
  M^\star(d, \epsilon) \triangleq \min\{M : \exists (M, d, \epsilon) -\text{code}\}. 
\end{align} 
Notably, when \( d = 0 \) and \( d(s, z) = \mathbf{1}\{s \neq z\} \), this corresponds to almost-lossless compression.  

In the fixed-to-fixed (block) setting, where \( \mathcal{S}^{k} \) and \( \mathcal{Z}^{k} \) are the \( k \)-fold Cartesian products of alphabets \( \mathcal{S} \) and \( \mathcal{Z} \), an \( (M, d, \epsilon) \) code for \( \{\mathcal{S}^k, \mathcal{Z}^k, P_{S^k}, d^k\} \) is referred to as a \( (k, M, d, \epsilon) \) code. For fixed \( \epsilon \), \( d \), and blocklength \( k \), the minimum achievable code size \( M^\star(k, d, \epsilon) \) and the finite blocklength rate-distortion function \( R^\star(k, d, \epsilon) \) are defined as
\begin{align}
  M^\star(k, d, \epsilon) \triangleq \min\{M : \exists (k, M, d, \epsilon)-\text{code}\},  
\end{align} 
\begin{align}
  R^\star(k, d, \epsilon) \triangleq \frac{1}{k} \log M^\star(k, d, \epsilon).  
\end{align} 

For variable-length coding, a code consists of mappings \( \mathsf{f}: \mathcal{S} \mapsto \{0, 1\}^* \) and \( \mathsf{g}: \{0, 1\}^* \mapsto \mathcal{Z} \), where \( \{0, 1\}^* \) denotes the set of all binary strings. Such a code operates at distortion level \( d \) if \( \mathbb{P}[\mathsf{d}(S, \mathsf{g}(\mathsf{f}(S))) \leq d] = 1 \). For a code \( (\mathsf{f}, \mathsf{g}) \) operating at distortion \( d \), the length of the binary codeword assigned to \( s \in \mathcal{S} \) is denoted by \( \ell(\mathsf{f}(s)) = \text{length of } \mathsf{f}(s) \).

\subsection{Lossless Compression}
\label{subsec:lossless sc}
Lossless data compression can be divided into two settings based on whether the code length is fixed. One setting is \textit{almost lossless fixed-length data compression} while the other is \textit{strictly lossless variable-length data compression}~\cite{kontoyiannis2013optimal}. Variable-length lossless compression is classified further by the use of prefixes. A reasonable way to characterize the performance of fixed-to-variable codes is to use their average encoded length. \(R^\star(k,\epsilon)\) is the minimum rate such that the probability that the best code’s compression rate is above \(R\) bits per symbol is no more than \(\epsilon\), i.e.,
\begin{align} \label{eq:lossless_R_opt}
\min_{\mathsf{f}} \mathbb{P}[\ell(\mathsf{f}(S^k)) > kR] \leq \epsilon.
\end{align}  
For prefix coding, the minimization should be performed under the prefix condition. Prefix coding requires that no codeword is a prefix of any other codeword. This property ensures that a long stream of fixed-to-variable length encoded symbols can be parsed unambiguously and decoded instantly.\footnote{For single-block compression, where the start and end are known, the prefix condition is less critical.}

Another fundamental limit at finite blocklengths is \( \epsilon^\star(k, M) \), which gives the best achievable excess-rate probability
\begin{align} \label{eq:lossless_Pe_opt}
\epsilon^\star(k, M) \triangleq \min_{\mathsf{f}} \mathbb{P}[\ell(\mathsf{f}(S^k)) \geq \log M].
\end{align}  
Herein the error event occurs when the length of the compressed codeword exceeds \( \log M \), where \( M \) is the number of distinct outcomes produced by the compressor.   
Verdú in~\cite{verdu2011teaching} and Kontoyiannis et al. in~\cite{kontoyiannis2013optimal} showed that the fundamental limit for strictly lossless variable-length codes without the prefix constraint, \(\epsilon^\star(k, M)\), equals the minimal error probability in fixed-length almost lossless codes. This result holds in both the nontrivial compression case where \(M<|\mathcal{S}|^k\) and the trivial case. The trivial case is omitted for brevity. In the nontrivial case, the optimal fixed-to-fixed compressor assigns a unique binary string of length \( \log M \) to each of the \( M - 1 \) most probable elements from \(\mathcal{S}^k\). It then assigns the remaining elements to another binary string of length \( \log M \), which indicates a coding failure. Only the strings encoded with lengths less than \( \log M \) by the optimal code can be decoded without error.
  
\subsubsection{Nonasymptotic Bounds}

Earlier, we considered sequences' notation, incorporating the blocklength \( k \). In presenting general nonasymptotic bounds, we apply a single-shot notation for clarity and simplicity, as in \cite{kostina2013lossy, Yury_phd}. Specifically, the random variable \( S \) and its realization \( s \) are abstract symbols that can be used to the entire sequences of length \( k \). When representing sequences, the corresponding alphabet \( \mathcal{S} \) should be interpreted as the \( k \)-fold Cartesian product of the single-letter alphabet.

The information of a random source output \( S \) with distribution \(P_{S}\) is defined according to~\cite[eq. (3)]{kontoyiannis2013optimal}
\begin{align} \label{eq:lossless_information}
  \imath_S(s)=\log\frac{1}{P_S(s)}.
\end{align} 
Not only is the distribution of the optimal code lengths \( \ell(\mathsf{f}^\star(S))\) closely linked with the distribution of \(\imath_S(S)\), where \( \mathsf{f}^\star\) denotes an optimal compressor, it is also important to note that similar information random variables play a crucial role in obtaining the fundamental limits of nonasymptotic information theory.
  
1)~Achievability Bounds: 
There are two main methods that are used to analyze the achievable bounds for the best codes. One method is by analyzing the information random variable \(\imath_S(S)\) defined in~\eqref{eq:lossless_information} to yield a bound on the code length produced by the optimal encoder. The other method is by exactly analyzing random binning to derive a lower bound for the performance of the optimal code.
For any source with a finite or countably infinite alphabet, a simple and powerful achievable bound states that the optimal encoder produces a code length that does not exceed the inherent information of the source. In other words, the distribution function of \( \ell(\mathsf{f}^\star(S)) \) dominates that of \( \imath_S(S) \).
This result was further refined in~\cite{verdu2007teaching} to bound the tail probabilities of both quantities, i.e., \( \mathbb{P}[\ell(\mathsf{f}^\star(S))\ge a] \le \mathbb{P}[\imath_S(S)\ge a] \). The observation comes from arranging the elements of \(\mathcal{S}\) in non-increasing order. Then, the probability of each element is upper bounded by the reciprocal of its rank in this order. 
Analogous to random coding in channel coding~\cite{polyanskiy2010channel}, one method for achievable error probability at finite blocklengths is \textit{random binning} in~\cite{kontoyiannis2013optimal}. In the work on lossy compression such as~\cite{kostina2012fixed}, \textit{random coding} terminology was used directly. In this setting, the compressor is no longer required to be an injective mapping. When the decompressor receives a label that can be explained by more than one source realization, it chooses the most likely one, breaking ties arbitrarily. This approach also introduces new computational challenge, especially for large blocklengths.

2)~Converse Bounds: 
Some works are based on the information variable to derive converse bounds, which concern the codeword lengths output by the optimal encoder \( \mathsf{f}^\star \).
Verdú in~\cite{verdu2007teaching} gave a converse bound of \(\ell(\mathsf{f}^\star(S))\),  
\(\max_{\tau>0}\left[\mathbb{P}[\imath_S(S)\ge \log M+\tau]-2^{-\tau}\right]\le\mathbb{P}[\ell(\mathsf{f}^\star(S))\ge \log M]\).  
This bound was obtained by considering a subset of the source alphabet \(\mathcal{L}=\{i\in \mathcal{S}:P_S(i)\le2^{-\tau}M^{-1}\}\) for a fixed arbitrary \(\tau>0\). Later, Kontoyiannis et al. in~\cite{kontoyiannis2013optimal} compared the code lengths \(\ell(\mathsf{f}(S))\) of an arbitrary compressor with the information random variable \(\imath_S(S)\). 
This result greatly advances pointwise asymptotic results and leads to the conclusion that the source dispersion of a source \(\{P_{S^{k}}\}_{k=1}^{\infty}\), \(\limsup_{k \to \infty} \frac{1}{k} \mathrm{Var}[\ell(\mathsf{f}_{k}^\star(S^{k}))]\)  is equal to its varentropy (minimal coding variance), \(\mathrm{Var}[\imath_S(S)]\).

Although the method based on the information random variable has been widely used for obtaining nonasymptotic limits, it does not always provide a tight bound~\cite[Fig. 1]{kontoyiannis2013optimal}. This has prompted the development of alternative approaches such as random coding~\cite{kontoyiannis2013optimal,kostina2012fixed}, which we already introduced in achievability bound, and hypothesis testing~\cite{kostina2012fixed} when deriving the converse bound for the best lossy compression code.

\subsubsection{Approximations}

For prefix variable-length codes, the asymptotic behavior of the minimal average compression rate \(\bar{R}(k)\) was given in~\cite{cover2006elements} as a widely known result. The term average refers to the overall performance of all compressors and thus the rate is unrelated to the excess-rate probability. 
Kontoyiannis in~\cite{kontoyiannis1997second} later provided a different kind of Gaussian approximation for the length of any prefix code. Specifically, Kontoyiannis first bounded \( \ell(\mathsf{f}(S))\) by a random variable with an approximate Gaussian distribution, and then he sharpened this bound to a law of the iterated logarithm (LIL).
For the large deviations in the distribution of code lengths, Merhav showed in~\cite{merhav1991universal} that for some sources with memory, the prefix constraint and compressor universality do not lower the optimal error exponent based on large deviations analysis. The Lempel-Ziv compressor achieves that exponent~\cite{welch1984technique}. Then, Kontoyiannis in~\cite[Sec. III]{kontoyiannis2013optimal} fully described the asymptotic behavior of the minimal average compression rate in terms of the entropy rate \(H(S)\).

Based on the close correspondence between optimal almost-lossless fixed-to-fixed codes and optimal strictly lossless fixed-to-variable codes, the following works apply to both settings. In the asymptotic regime, Csiszár et al. in~\cite{csiszar2011information} parameterized the exponential decrease of the error probability. Szpankowski et al. in~\cite{szpankowski2011minimum} provided an approximation for the minimal average compression rate for non-equiprobable sources. Unlike prefix codes in~\cite{cover2006elements}, an extra second-order term \(-\frac{2}{k}\log k\) was obtained for the approximation of \(\bar{R}(k)\). This result was later extended to the case where \(\imath_S(S)\) is non-lattice\footnote{A discrete random variable is lattice if all its masses lie on a subset of some lattice \(\{\nu+n\zeta\}\) with \(n\in\mathbb{Z}\).}. On the other hand, for the minimum achievable source coding rate \(R^{\star}(k,\epsilon)\), Yushkevich in~\cite{yushkevich1953limit} derived an approximation. Strassen in~\cite{strassen1962asymptotische} extended this result to non-equiprobable memoryless sources such that \(\imath_S(S)\) is non-lattice. Kontoyiannis in~\cite{kontoyiannis2013optimal} argued that Strassen's complete proof of the approximation for \(R^{\star}(k,\epsilon)\) is controversial, and he provided
\begin{align}
R^{\star}(k,\epsilon)=H(S)+\sqrt{\frac{V(S)}{k}}\,Q^{-1}(\epsilon)-\frac{1}{2k}\log k+O\left(\frac{1}{k}\right),
\end{align}
and corresponding detailed proof. The two key quantities are defined as \(H(S) = \mathbb{E}[\imath_S(S)]\) and \(V(S) = \mathrm{Var}[\imath_S(S)]\). \(Q^{-1}(\cdot)\) denotes the inverse of the complementary cumulative distribution function (CDF) of the standard Gaussian distribution. Intuitively, this occurs because by the central limit theorem the distribution of \( \imath_{S^k}(S^k)=\sum_{i=1}^k \imath_S(S_i) \) is approximately Gaussian, where \( S^k \) denotes the source output sequence of length \( k \). The result is obtained through the application of precise converse and achievability bounds together with the classical Berry-Esséen bound~\cite{prokhorov2000limit}.

\subsection{Lossy Compression}
\label{subsec:lossy sc}
The core problem of lossy data compression is to represent an object under a compression rate constraint while meeting a reproduction criterion. In channel coding or almost lossless compression, block error rates serve as the performance metric. In contrast, lossy compression can be evaluated using symbol error rates~\cite{kostina2013lossy}.

\subsubsection{Nonasymptotic Bounds}

Inspired by the information random variable defined for lossless compression in~\eqref{eq:lossless_information}, an important quantity in nonasymptotic theory for lossy compression is the \(\mathsf{d}\)-tilted information~\cite{kostina2012fixed}. Based on it, one can obtain both the converse bounds and achievability bounds. We first introduce it, i.e.,
\begin{align} \label{eq:d_tilted_information}
  \jmath_S(s,d) = \log \frac{1}{\mathbb{E}\left[\exp\left\{\lambda^\star d - \lambda^\star \mathsf{d}(s, Z^\star)\right\}\right]},
\end{align}
which essentially quantifies the number of bits needed to represent the source output \(s\) within distortion \(d\). The Lagrange multiplier is given by \( \lambda^\star = -\mathbb{R}_S'(d) \), and the infomation rate-distortion function is given by \( \mathbb{R}_S(d) = \mathbb{E}[\jmath_S(S,d)] \). When \(d = 0\) and for discrete random variables with \(d(s, z) = \mathbf{1}\{s \neq z\}\), it is natural to define the 0-tilted information as \(\jmath_S(s, 0)\), which reduces to the information random variable \(\imath_S(s)\) in~\eqref{eq:lossless_information} for the almost lossless case. Furthermore, the average value of \( \jmath_{S^k}(S^k,d) \) equals the asymptotic optimal rate \( kR(d) \) by the intuition that long sequences tend to approach their mean\footnote{For simplicity, we omit the superscript on the minimal achievable coding rate in the following.}. Here, the asymptotic fundamental limit \(R(d)\) is the supremum of \(R(k,d,\epsilon)\) as the blocklength \(k\) tends to infinity, i.e. \(R(d)=\sup_{k\to\infty} R(k,d,\epsilon)\). \(R(k,d,\epsilon)\) is the coding rate corresponding to the minimum codebook size \(M^\star(k,d,\epsilon)\) in the nonasymptotic regime.
Tight nonasymptotic bounds relate the probability that a code with \(M\) representation points yields distortion above \(d\) (operational quantity) to the probability that the \(\mathsf{d}\)-tilted information exceeds \(\log M\) (information-theoretic quantity). These two quantities mirror the classification in the asymptotic regime, where achievable rate-distortion pairs are defined from an operational perspective and the rate-distortion function from an informational perspective.

1)~Converse Bounds: 
Shannon established the fundamental rate‐distortion limits for coding with average distortion in~\cite{shannon1959coding} . Later, Körner et al. in~\cite{korner1971coding} and Kieffer in~\cite{kieffer1991strong} proved a strong converse bound for lossy source coding, indicating that if a fixed compression rate \(R\) satisfies \(R < \mathbb{R}_S(d)\), then the error probability \(\epsilon\) tends to one as \(k \to \infty\). For prefix‐free variable‐length lossy compression, a key nonasymptotic converse bound was derived by Kontoyiannis in~\cite{kontoyiannis2000pointwise}. For a discrete memoryless source with the finite alphabet and a bounded separable distortion measure, one can obtain a finite blocklength converse bound from Marton’s fixed‐rate error exponent in~\cite{marton1974error}. Later, Han in~\cite{han2006reliability} and Iriyama in~\cite{iriyama2001probability} extended the error exponent analysis method to obtain nonasymptotic theoretical results.
In addition to obtaining a converse bound using the \(\mathsf{d}\)-tilted information, inspiration from \cite{polyanskiy2010channel} led to a potentially tighter bound in certain cases based on binary hypothesis testing~\cite{kostina2012fixed}. 
Let \( Q \) be an auxiliary distribution defined on the alphabet \( \mathcal{S} \). Consider any randomized test \( P_{W|X} : \mathcal{S} \mapsto \{0, 1\} \) where the output \(1\) favors the true source distribution \( P_S \). This approach leads to a lower bound on the size \( M \) of any code satisfying a given fidelity criterion,\footnote{\(P\) and \(Q\) denote the distributions, and \(\mathbb{P}\) and \(\mathbb{Q}\) represent event probabilities in the underlying space.} 
\begin{align} \label{eq:hypothesis_test_converse}
  M \geq \sup_Q \inf_{z \in \mathcal{Z}} \frac{\beta_{1-\epsilon}(P_S, Q)}{\mathbb{Q}\left[\mathsf{d}(S,z)\leq d\right]},
\end{align} 
For an observed source output \( s \in \mathcal{S} \), the optimal performance of binary hypothesis testing is defined as  
\(
\beta_\alpha(P, Q) \triangleq \min_{P_{W|X} : \mathbb{P}[W=1] \geq \alpha} \mathbb{Q}[W = 1]  
\).
Suppose the source \(S\) takes values on a countable alphabet \(\mathcal{S}\) and let the distribution \(Q\) be uniform on \(\mathcal{S}\). This choice yields a looser lower bound in~\eqref{eq:hypothesis_test_converse} but helps to better understand the bound. 
Consider a set \( \Omega \subset \mathcal{S} \) with a probability measure of \( 1 - \epsilon \). For any \( s \in \Omega \), the optimal binary hypothesis test with error probability \( \epsilon \) will choose \( P_S \) over \( Q \). Therefore, the type II error \( \beta_{1-\epsilon}(P_S, Q) \) is proportional to the number of elements in \( \Omega \), while \( \mathbb{Q}\left[\mathsf{d}(S, z) \leq d\right] \) is proportional to the number of elements that can be placed into a distortion ball of radius \( d \). Thus, the ratio leads to a lower bound on the minimum number of distortion balls needed to cover the set \( \Omega \). This lower bound is often not achievable due to the overlap between distortion \( \mathsf{d} \)-balls.

2)~Achievability Bounds: 
The most general achievability bound, which guarantees the existence of a code with an upper bound on the error probability, originates from Shannon in~\cite{shannon1959coding} and was later distilled by Verdú in~\cite{verdu2009ele528}. For three specific setups with independent and identically distributed (i.i.d.) sources and separable distortion measures, Goblick provided achievability bounds for fixed-rate compression of a finite alphabet source in~\cite{goblick1963coding}, Pinkston for variable-rate compression of a finite alphabet source in~\cite{pinkston1967encoding}, and Sakrison for variable-rate compression of a Gaussian source with mean-square error distortion in~\cite{sakrison1968geometric}. However, these bounds are often cumbersome to a certain extent. Later, Kostina et al. in~\cite{kostina2012fixed} developed two main approaches for obtaining achievability bounds, which were inspired by the methods used to analyze the non-asymptotic fundamental limits of lossless compression that we introduced earlier.
One approach encodes the source output using random coding. Here the excess-distortion probability comes from the event that none of the \(M\) codewords falls into the distortion ball of radius \(d\) centered at the source output \(s\). In fact, this approach is somewhat difficult to evaluate numerically because of its high computational complexity. An alternative method applies the (generalized) \( \mathsf{d} \)-tilted information to derive a lower bound on the probability that a codeword falls within the distortion ball around \(s\). In this method, a parameter \( \gamma \) is introduced to account for the “radius” of a spherical shell on the surface of the distortion \( \mathsf{d} \)-ball. This technique proves particularly useful when analyzing the boundaries of random coding methods.

\subsubsection{Approximations}
In variable-rate quantization, the lossy asymptotic equipartition property (AEP) yields strong achievability and converse bounds and is concerned with the asymptotic behavior of distortion \( \mathsf{d}\)-balls. Second-order refinements of the lossy AEP were investigated by Yang et al. in~\cite{yang1999redundancy} and Kontoyiannis in~\cite{kontoyiannis2000pointwise}. Later, an asymptotic approximation for the minimum achievable rate of sources on an arbitrary alphabet under fairly general conditions was derived by Kostina et al. in~\cite{kostina2012fixed}. This result was obtained through the application of tight nonasymptotic bounds and nonasymptotic refinements of the lossy AEP.

Before introducing the Gaussian approximation in~\cite{kostina2012fixed}, certain conditions need to be satisfied. First, the source \(\{S_i\}\) is assumed to be stationary and memoryless, and the distortion measure is required to be separable with an appropriately bounded distortion level. In addition, the ninth moment of the random variable \(\mathsf{d}(S, Z^\star)\) should be finite, where \(Z^\star\) is the reconstruction that achieves the rate-distortion function. Under these conditions, the minimum achievable rate satisfies
\begin{align} \label{eq:SC_min_approx_rate}
  R(k,d,\epsilon) &= R(d) + \sqrt{\frac{\mathcal{V}(d)}{k}}\, Q^{-1}(\epsilon) + \theta\left(\frac{\log k}{k}\right),
\end{align} 
where first-order term \(R(d)\) and the second-order term \(\mathcal{V}(d)\) are, respectively, the mean and variance of the \(\mathsf{d}\)-tilted information, i.e., \(R(d)=\mathbb{E}[\jmath_\mathsf{S}(\mathsf{S},d)]\), \(\mathcal{V}(d)=\mathrm{Var}[\jmath_\mathsf{S}(\mathsf{S},d)]\). 
The remainder term \(\theta\left(\frac{\log k}{k}\right)\) satisfies
\begin{align} \label{eq:SC_min_approx_rate_reminder}
  -\frac{1}{2}\frac{\log k}{k} + O\left(\frac{1}{k}\right) \le \theta\left(\frac{\log k}{k}\right) \le C_0\frac{\log k}{k} + \frac{\log\log k}{k} + O\left(\frac{1}{k}\right),
\end{align} 
where \(C_0 = \frac{1}{2} + \frac{\operatorname{Var}\!\left(J^{\prime}_{\mathsf{Z}^{\star}}(\mathsf{S},\lambda^\star)\right)}{\mathbb{E}\!\left[\lvert J^{\prime\prime}_{\mathsf{Z}^{\star}}(\mathsf{S},\lambda^\star) \rvert\right]\log e}\) with \(\lambda^\star = -\mathbb{R}_S'(d)\). The approximation in \eqref{eq:SC_min_approx_rate} quantitatively characterizes how the minimum compression rate deviates from the rate-distortion function in the finite blocklength regime. The first term corresponds to the asymptotic RDF, while the remaining terms can be interpreted as penalty terms arising from the finite blocklength constraint. In particular, the second-order term is of primary interest, as it explicitly depends on the variance of the \(\mathsf{d}\)-tilted information and the excess distortion probability threshold \(\epsilon\), with the Gaussian approximation yielding the \(Q^{-1}(\cdot)\) function. As for the remainder term in \eqref{eq:SC_min_approx_rate_reminder}, its order with respect to \(k\) is established, but obtaining sharper bounds on it remains an open problem.

These general results are applied to binary memoryless sources (BMS), discrete memoryless sources (DMS), and Gaussian memoryless sources (GMS), accompanied by simulation results for error probability and codebook size bounds, as well as approximations of the minimum achievable compression rate. For example, for GMS, the expression of the \(\mathsf{d}\)-tilted information \(\jmath_{S^k}(s^k,d)\) can be derived, where \(S^k \sim \mathcal{N}(0,\sigma^2)\) denotes a length-\(k\) i.i.d. Gaussian source, and \(Z^{k\star} \sim \mathcal{N}(0,\sigma^2-d)\) represents the optimal reproduction source. For the converse based on hypothesis testing, a commonly used and effective choice of the auxiliary distribution \(\mathbb{Q}\) for Gaussian sources is the Lebesgue measure. Regarding achievability via random coding, the spherical symmetry of Gaussian sources makes random coding closely related to the classical sphere covering problem. As for the approximation part, the second-order variance term \(\mathcal{V}(d)\) can be readily obtained, since the distribution of \(\jmath_{S^k}(s^k,d)\) is explicit. For further details on the remainder term bounds, readers are referred to~\cite{kostina2012fixed,kostina2013lossy}.

\subsubsection{Gauss-Markov Source}
We have introduced memoryless sources earlier~\cite{kostina2012fixed,shannon1959coding}. Yet many practical sources, such as images and videos, have memory and exhibit correlations at both pixel and frame levels~\cite{varodayan2006exploiting, memon1996lossless}. Therefore, it is important to extend research to sources with memory. Compared to memoryless sources, most tight results for sources with memory are initially limited and appear only in the asymptotic regime~\cite{zhou2023finite}. Kolmogorov first derived the rate-distortion function for stationary Gaussian autoregressive sources under the quadratic distortion measure in~\cite{kolmogorov1956shannon}. In~\cite{berger1970information}, Berger then generalized this result to a non-stationary case, i.e., the Wiener process. Gray later extended it to the general Gaussian autoregressive source and first-order binary symmetric Markov processes in~\cite{gray1970information}. 
A common model for sources with memory is a Gaussian source with first-order Markovian memory~\cite{davisson1972rate}. This is known as a Gauss-Markov source and is a special case of the Gaussian autoregressive source~\cite{gray1970information}. 
The main progress in nonasymptotic fundamental limits for sources with memory was made by Tian et al. They analyzed stationary Gauss-Markov sources in~\cite{tian2019dispersion} and non-stationary cases in~\cite{tian2021nonstationary}. Here, we take the stationary case as an example and introduce the key methods used to obtain nonasymptotic fundamental limits for Gauss-Markov sources, by presenting the derivation of nonasymptotic achievability, converse bounds and the approximations.

A Gauss-Markov source can be modeled as, for \(\forall i \ge 1\), 
\begin{align}
  U_i = aU_{i-1} + Z_i,
\end{align}  
with \(U_0=0\), and where the \(Z_i\) are i.i.d. Gaussian random variables with \(Z_i \sim \mathcal{N}(0,\sigma^2)\). For \(|a|<1\), \(|a|>1\), and \(|a|=1\), they correspond to stationary sources, non-stationary sources, and the Wiener process, respectively. 
In~\cite{tian2019dispersion}, the \(\mathsf{d}\)-tilted information \(\jmath_{X^k}(x^k,d)\) originally defined for a GMS in~\cite{kostina2012fixed} was extended to the Gauss-Markov source, inherently incorporating the reverse waterfilling principle. 

1)~Converse Bounds: 
There are two main methods for obtaining converse bounds. One based on the volumetric argument and the other on the \( \mathsf{d} \)-tilted information. Thanks to the spherical symmetry of Gaussian distributions, the volumetric argument arises. In the codeword space, the distortion \( \mathsf{d} \)-ball is stretched or compressed along certain axes due to correlations, possibly transforming it into an ellipsoid. However, the overall volume remains unchanged because correlation does not alter the source's energy. Thus, the converse bound is essentially the same as that for the i.i.d. Gaussian case. Unlike the i.i.d. Gaussian case, the volumetric method can only yield the optimal second-order coding rate for the Gauss-Markov source in the low-distortion regime~\cite[Theorem 7]{tian2019dispersion}. A more general approach still relies on the \( \mathsf{d} \)-tilted information of the decorrelated sources, please refer to~\cite{tian2019dispersion}.

2)~Achievability Bounds: 
Dumer et al. in~\cite{dumer2004coverings} examined the problem of covering an ellipsoid in \(\mathbb{R}^k\) with the minimum number of balls and derived both lower and upper bounds on the covering number. Inspired by sphere covering techniques, however, applying their result in~\cite{dumer2004coverings} directly to the achievability bounds for the minimum achievable rate of Gauss-Markov sources yields very loose results. This is because the loss of spherical uniformity and the linear transformation (or decorrelation) of the distortion \(\mathsf{d}\)-ball affect the covering number of the ellipsoid drastically. A more reliable approach for the achievability bound is to construct a typical set based on the maximum likelihood estimator, which relies on the lossy AEP for Gauss-Markov sources~\cite{tian2019dispersion}.

3)~Approximations:
The minimum achievable rate for the Gauss-Markov source satisfies~\cite[Theorem 1]{tian2019dispersion}
\begin{align}
  R(k, d, \epsilon) = \mathbb{R}_U(d) + \sqrt{\frac{V_U(d)}{k}}\, Q^{-1}(\epsilon) + o\!\left(\sqrt{\frac{1}{k}}\right),
\end{align} 
where \(\mathbb{R}_U(d)\) is the rate-distortion function. In the second-order term, the operational dispersion is given by
\begin{align}
  V_U(d) = \frac{1}{4\pi} \int_{-\pi}^{\pi} \min\left\{1, \left(\frac{S(w)}{\theta}\right)^2\right\}\, dw,
\end{align} 
where \(\theta > 0\) is the water level corresponding to the distortion \(d\), and the power spectrum is defined as \(S(w) = \frac{\sigma^2}{g(w)}\) with \(g(w) = 1+a^{2}-2a\cos(w)\).

\subsubsection{Mismatch}
The mismatch problem arises from a practical challenge in lossy data compression where a codebook is designed for a source with one distribution but used to compress a different source with another distribution. In other words, the source to be compressed is not matched to the pre-designed codebook. For an arbitrary memoryless source under the quadratic distortion measure, Lapidoth used a spherical codebook and minimum Euclidean distance encoding to compress the source in~\cite{lapidoth1997role}. 
Lapidoth concluded that for any ergodic source with a known finite second moment \( \sigma^2 \), the rate-distortion function for the GMS with a distribution \( \mathcal{N}(0,\sigma^2) \) is achievable and ensemble tight as the blocklength increases. Ensemble tight means the code analysis is optimal. It is worth noting that Lapidoth’s work advanced the solution to the mismatch problem because only the source second moment is required. This quantity is easier to obtain than the full source distribution and can be estimated from an observed source sequence. Later it was extended to two codebook types by Zhou et al. in~\cite{zhou2018refined}. One codebook is spherical. Every codeword is generated independently and uniformly on the surface of a sphere with radius \( \sqrt{k(\sigma^2-d)} \). The other codebook is Gaussian. Every codeword is drawn from a product Gaussian distribution with zero mean and variance \( \sigma^2-d \). Unlike the case with a known source distribution, the performance is evaluated by the ensemble excess-distortion probability evaluated with respect to both the source and the codebook distributions. Zhou et al. in~\cite{zhou2018refined} improved Lapidoth's first-order asymptotic result by deriving a second-order approximation for the minimum achievable coding rate. They extended Lapidoth’s work to consider two different types of codebooks and concluded that both achieve the same first-order and second-order optimality.

For any memoryless source \(S\) satisfying \(\mathbb{E}[S^2]=\sigma^2\), \(\zeta=\mathbb{E}[S^4]<\infty\), and \(\mathbb{E}[S^6]<\infty\), the second-order approximation for a given codebook type \(\dagger \in \{\mathrm{sp, iid}\}\) is given by~\cite[Theorem 1]{zhou2018refined}
\begin{align} \label{eq:mismatched_rate}
  R^{\dagger}(k,\sigma^2,d,\epsilon) = \log\left(\frac{\sigma^2}{d}\right) + \sqrt{\frac{V(\sigma^2,\zeta)}{k}} \, Q^{-1}(\epsilon) + O\!\left(\frac{\log k}{k}\right),
\end{align} 
where the mismatched dispersion is defined as \(V(\sigma^2,\zeta) \triangleq \frac{\zeta-\sigma^4}{4\sigma^4}\).


Intuitively, the primary error event arises from the atypicality of the source sequence with an unknown distribution. However, regardless of which of the two codebook types is used, roughly \(\exp\!\left(\frac{k}{2}\log\frac{\sigma^2}{d}\right)\) codewords suffice to cover all typical sequences with an error probability that decays faster than exponentially. This result was obtained through a careful analysis of the probability of atypical source sequences. For more details, please refer to~\cite{zhou2018refined}.

\subsection{Multiterminal Setting}
\label{subsec:multiterminal sc}
Attention now shifts to multiterminal settings. We focus on two main problems. In the first problem, two decoders recover the output of a common encoder. Side information is available at both the encoder and one decoder. This problem was first introduced by Kaspi in~\cite{kaspi2002rate}, also known as the ``Kaspi problem''.  In the second problem, an encoder-decoder pair is added to the classical rate-distortion formulation. This setup forms a system with two encoders and two decoders. It is known as the successive refinement problem~\cite{equitz1991successive}.

\subsubsection{Kaspi Problem}
\begin{figure}[t]
  \centering
  \includegraphics[width=0.6\linewidth]{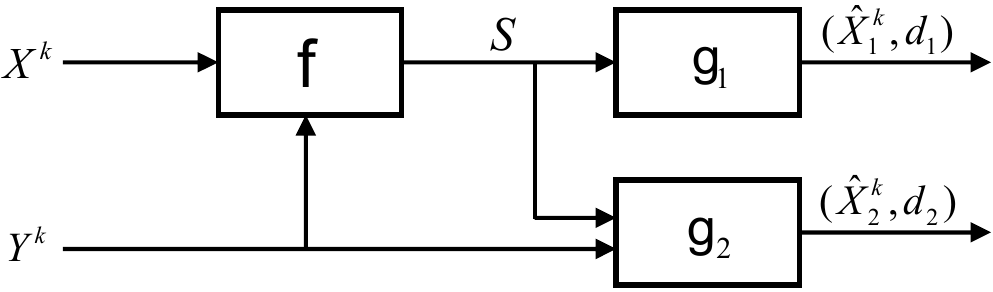}
  \caption{System model for the Kaspi problem.} \label{Fig:kaspi}
\end{figure}

The Kaspi problem model is shown in Fig.~\ref{Fig:kaspi}, where a common encoder \( \mathsf{f}\) and two decoders \( \mathsf{g}_i\) for \(i=1,2\) are employed. The side information \(Y^k\) is available only at \(\mathsf{g}_2\). Since there are two decoder outputs \(\widehat{X}_i^k\) for \(i=1,2\), two different distortion measures and corresponding constraints are used. In the nonasymptotic regime the performance is measured by the joint excess-distortion probability \(P_{e,k}(d_1,d_2) \triangleq \mathbb{P}\left[\mathsf{d}_1(X^k,\widehat{X}_1^k)>d_1 \text{ or } \mathsf{d}_2(X^k,\widehat{X}_2^k)>d_2\right]\), which takes into account both the source and the codebook distributions. 
In fact, separate excess-distortion probabilities were used to assess the performance of optimal codes in~\cite{no2016strong}. However, Zhou et al. in~\cite{zhou2017second} pointed out that the joint excess-distortion probability offers several advantages over the separate ones. Therefore, we focus on results based on the joint excess-distortion probability criterion.

The point-to-point \(\mathsf{d}\)-tilted information was extended to the \((\mathsf{d}_1,\mathsf{d}_2)\)-tilted information for the Kaspi problem in~\cite{zhou2019non}. One key property of \((\mathsf{d}_1,\mathsf{d}_2)\)-tilted information is that the expectation with respect to the source and side infomation equals the rate-distortion function, i.e. \(R(P_{XY},d_1,d_2)=\mathbb{E}[\jmath_K(X,Y\mid d_1,d_2,P_{XY})]\). Based on the property of \((\mathsf{d}_1,\mathsf{d}_2)\)-tilted information and Kostina et al.'s one-shot converse argument in~\cite{kostina2012new}, Zhou et al. in~\cite[Lemma 5]{zhou2019non} derived a converse bound for the Kaspi problem. In the achievability part, a type covering lemma for the Kaspi problem was obtained and the properties of the \((\mathsf{d}_1,\mathsf{d}_2)\)-tilted information were used with an appropriate Taylor expansion. In addition to applying the Berry-Esséen Theorem to derive an asymptotic approximation, Zhou et al. in~\cite[Sec. III, Sebsec. D]{zhou2019non} used the large and moderate deviations to derive the asymptotics of the error exponent for DMSes. Here, we mainly introduce the optimal second-order coding rate for the Kaspi problem through the Berry-Esséen Theorem.

Let \(V(d_1,d_2,P_{XY})\) denote the distortion-dispersion function for the Kaspi problem, that is,  
\(V(d_1,d_2,P_{XY}) = \operatorname{Var}\left[\jmath_K(X,Y\mid d_1,d_2,P_{XY})\right]\). A rate \(L\) is said to be second-order \((d_1,d_2,\epsilon)\)-achievable if there exists a sequence of \((k,M)\)-codes such that \(\limsup_{k\to\infty}\frac{1}{\sqrt{k}}(\log M - k\,R(P_{XY},d_1,d_2)) \le L\) and \(\limsup_{k\to\infty}P_{e,k}(d_1,d_2) \le \epsilon\). The optimal second-order coding rate is defined as the infimum over all such achievable rates and is denoted by \(L^\star(d_1,d_2,\epsilon)\), which is given by 
\begin{align}
  L^\star(d_1,d_2,\epsilon) = \sqrt{V(d_1,d_2,P_{XY})}\,Q^{-1}(\epsilon).
\end{align} 
It is worth mentioning that for different distortion levels \((d_1,d_2)\), the key quantity \(\jmath_K(x,y\mid d_1,d_2,P_{XY})\) for the Kaspi problem reduces to that of other cases. For example, when decoder \(\mathsf{g}_2\) is removed, the Kaspi problem reduces to the conventional lossy source coding problem and \(\jmath_K(x,y\mid d_1,d_2,P_{XY})\) reduces to the \(d_1\)-tilted information in~\cite{kostina2012fixed}. Similarly, when decoder \(\mathsf{g}_1\) is removed, the setup reduces to the case with side information available at both the encoder and decoder and \(\jmath_K(x,y\mid d_1,d_2,P_{XY})\) reduces to the \(d_2\)-tilted information in~\cite{le2014second}.

\subsubsection{Successive Refinement}
\begin{figure}[t]
  \centering
  \includegraphics[width=0.6\linewidth]{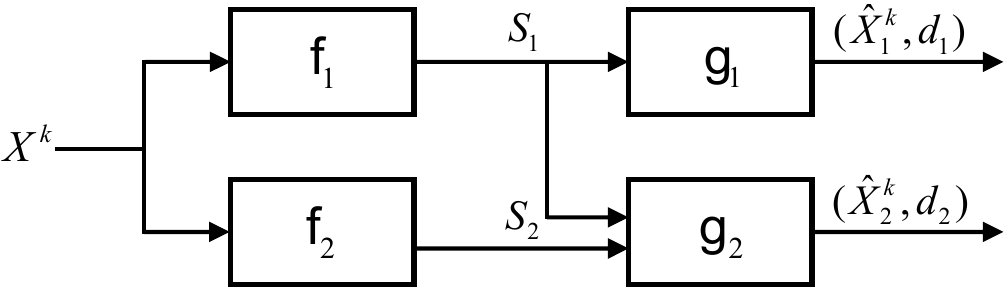}
  \caption{System model for the successive refinement problem.} \label{Fig:successive refinement}
\end{figure}
The successive refinement problem is illustrated in Fig.~\ref{Fig:successive refinement}. An additional decoder accesses the compressed outputs from both encoders simultaneously~\cite{equitz1991successive}. This additional decoder produces a more precise reconstruction of the source sequence than a model in which the decoder receives information from only one encoder. Moreover, the successive refinement formulation relates to the question of whether it is possible without sacrificing the optimality of lossy compression to interrupt a transmission and achieve a better reconstruction of certain source sequences~\cite{rimoldi1994successive}. For any distortion measure, Rimoldi in~\cite{rimoldi1994successive} characterized the optimal rate-distortion region for a DMS. Effros later generalized Rimoldi's work to discrete stationary ergodic and non-ergodic sources in~\cite{effros1999distortion}. Kanlis et al. in~\cite{kanlis1996error} obtained the error exponent under the joint excess-distortion criterion, while Tuncel et al. in~\cite{tuncel2003error} considered the separate excess-distortion criterion. No et al. in~\cite{no2016strong} derived the second-order coding rates for the strong successive refinement problem. Zhou et al. in~\cite{zhou2017second} derived the optimal second-order coding region and the moderate deviations constant for the successive refinement source coding problem under the joint excess-distortion criterion for a DMS with arbitrary distortion measures. They also considered a GMS with the quadratic distortion measure, where the results are especially simple since a GMS with the quadratic distortion measure is successively refinable~\cite{equitz1991successive}.



Considering a general DMS, let the rate-dispersion matrix \( \mathbf{V}(R_1^\star,d_1,d_2 \mid P_X) \succ 0 \) denote the covariance matrix of the two-dimensional random vector \([\,\jmath(X,d_1 \mid P_X),\, \jmath(X,R_1^\star,d_1,d_2 \mid P_X)\,]^T\). Under certain conditions, the second-order coding region was characterized in~\cite[Theorem 11]{zhou2017second} and divided into three cases. In the first two cases, the code has a rate bounded away from one of the first-order limits, so that the second-order behavior can be captured by a univariate Gaussian distribution. In contrast, in the third case, the code operates exactly at both first-order limits, which requires a bivariate Gaussian formulation to capture the second-order behavior. This result holds for both positive-definite and rank-deficient rate-dispersion matrices, following an argument by Tan et al. in~\cite[Theorem 6]{tan2013dispersions}.

\section{Channel Coding}
\label{sec:cc}
The basic task of channel coding is to transmit \(M\) messages over a noisy channel so that they can be distinguished reliably at the receiver. In this section, we review recent advances in the information-theoretic fundamental limits of channel coding, with a particular focus on P2P, multiple access, and massive access channels. 

\subsection{Point-to-Point Channels}
\label{subsec:pp cc}
\subsubsection{Key Metrics}

Shannon's foundational communication framework \cite{Shannon} formalizes channel coding through four essential components: 1)~an apriori unknown message, which is modeled as a random variable \(W\) equiprobable on the set \(\{1,\ldots,M\}\); 
2)~an encoder \(\mathsf{f}: \{1,\ldots,M\} \to \mathcal{X}^n\), which maps a message \(W\) into a codeword \(X^n\) of length \(n\); 
3)~a channel, which is modeled as a sequence of random transformations \(P_{Y^n|X^n}\) with input \(x^n \in \mathcal{X}^n\) and output \(y^n \in \mathcal{Y}^n\); 
and 4)~a decoder \(\mathsf{g}: \mathcal{Y}^n \to \{1,\ldots,M\}\) that outputs the estimated message \(\widehat{W}\) based on the channel output \(Y^n\).

An \((n,M,\epsilon)\)-code is defined as the encoder-decoder pair \((\mathsf{f}, \mathsf{g})\) with blocklength \(n\) and codebook size \(M\) guaranteeing that the decoding error probability is below \(\epsilon\). 
The commonly used two kinds of error constraints, i.e. the average error probability constraint and the maximum error probability constraint, are given by: 
\begin{equation}
    P_{{\rm e, ave}} = \mathbb{P} \left[ \widehat{W} \neq W  \right]  \leq \epsilon , \label{eqc:pe_ave}
\end{equation}
\begin{equation}
    P_{{\rm e, max}} = \max_{1\leq j\leq M} \mathbb{P} \left[ \widehat{W} \neq W | W = j \right]  \leq \epsilon. \label{eqc:pe_max}
\end{equation} 
The rate is defined as \(R \triangleq \frac{\log M}{n}\), which is measured in bits per channel use. 
The fundamental limit \(R^{\star}(n,\epsilon)\), representing the maximum data rate under blocklength \(n\) and target error probability \(\epsilon\), is defined as
\begin{equation}
    R^{\star}(n,\epsilon) \triangleq \sup  \left\{ R: \exists (n,M,\epsilon)-{\rm code} \right\}. \label{eqc:R}
\end{equation}
Likewise, the smallest achievable error probability is defined as
\begin{equation}
    \epsilon^{\star}(n,R) \triangleq \sup  \left\{ \epsilon : \exists (n, 2^{nR} ,\epsilon)-{\rm code} \right\}. \label{eqc:epsilon}
\end{equation}

\subsubsection{Classical Asymptotic Results}

Shannon's pioneering work \cite{Shannon} established the theoretical foundation for analyzing \(R^*(n,\epsilon)\) given an \((n, \epsilon)\) pair. 
A remarkable observation by Shannon was that as the blocklength \(n\) tends to infinity and the error probability \(\epsilon\) goes to \(0\), \(R^{\star}(n,\epsilon)\) becomes asymptotically tractable, and the asymptotic limit of \(R^{\star}(n,\epsilon)\) is known as the channel capacity \(C\), i.e., 
\begin{equation}
    C = \lim_{\epsilon \to 0} \lim_{n\to\infty} R^{\star}(n,\epsilon).
\end{equation}
It shows that error-free transmission remains feasible for any rate below capacity as long as the blocklength is sufficiently large. 
For a memoryless channel \(P_{Y|X}\), we can express the channel capacity \(C\) as \cite{Shannon}, \cite[Sec. 7]{cover2006elements} 
\begin{equation}
    C = \sup_{P_X} I(X;Y), 
\end{equation}
where \(I(X;Y)\) denotes the single-letter mutual information between \(X\) and \(Y\), and the supremum is taken over all input distributions. 
 For a fixed \(\epsilon>0\), as \(n\to\infty\), the limit of \(R^{\star}(n,\epsilon)\) is known as the $\epsilon$-capacity \(C_{\epsilon}\), i.e., 
 \begin{equation}
     C_{\epsilon} = \lim_{n\to\infty} R^{\star}(n,\epsilon).
 \end{equation}

For a fixed rate \(R\), the asymptotic behavior of \(\epsilon^{\star}(n,R)\) is determined by the reliability function \(E(R)\), i.e.,
\begin{equation} 
    E(R) = \liminf_{n\to\infty}  - \frac{\log \epsilon^{\star}(n,R)}{n}. 
\end{equation}
For any rate \(R\) exceeding the capacity \(C\), the communication is unreliable with \(E(R) = 0\). By restricting \(R < C\), the error probability is able to decay exponentially with the exponent \(E(R)>0\).



\subsubsection{Non-Asymptotic Bounds}

Classical information theory, which relies on the assumptions of infinite blocklength, infinite payload size, and/or vanishing error probability, fails to characterize the fundamental limits of practical communication systems that employ short blocklength and short packet and require small but nonnegligible error probability. 
To circumvent this problem, a series of works have focused on finite-blocklength information theory. 
In the finite-blocklength regime, exact computation of the maximal rate \(R^*(n,\epsilon)\) is computationally intractable, even for the simple binary symmetric channel and binary erasure channel, motivating researchers to develop tight and computationally tractable bounds on both achievability and converse sides. 
In the following, we review some non-asymptotic bounds and the key ideas used to derive these results. 

1)~Achievability Bounds: 
Feinstein~\cite{Feinstein} and Shannon~\cite{Shannon2} established finite-blocklength achievability bounds (i.e., lower bounds) on \(R^{\star}(n,\epsilon)\) based on the maximal coding and random coding ideas, respectively. 
According to the random coding idea, one needs to randomly construct the codebook from some distribution, and then evaluate the error probability across ensemble realizations, thereby proving the existence of codes under the constraint on the average error probability. 
Based on the maximal coding approach, one needs to sequentially append codewords until violating the constraint on the maximal probability of error. 
These two ideas have been applied in almost all achievability bounds in the literature. 
These bounds differ primarily in decoding rules, such as typicality decoding, maximal-likelihood decoding, and threshold decoding, and error analysis techniques.  
For instance, Polyanskiy, Poor, and Verd\'u \cite{polyanskiy2010channel} derived several analytically tractable and non-asymptotically sharp achievability bounds on \(R^{\star}(n,\epsilon)\), including the dependence-testing (DT) bound and the \(\kappa\beta\) bound. 
The key principles used to derive these achievability bounds are random/maximal coding and hypothesis testing. 
In addition to the bound on \(R^{\star}(n,\epsilon)\), Gallager~\cite{Gallager_1965} derived an achievability bound more suitable for the analysis of the reliability function (see \cite{Scarlett_phd} for a recent survey).
Moreover, the connections between the asymptotic golden formula and non-asymptotic $\beta\beta$ bounds were characterized in \cite{yang2018beta}. 
The non-asymptotic fundamental limits of wiretap channels were explored in \cite{yang2019wiretap}.

The above bounds can be naturally generalized to power-constrained systems. 
The input distribution critically influences the finite-blocklength performance. 
Concentrating the power of most codewords near the maximum available power budget substantially enhances the performance. 
Therefore, instead of using i.i.d. Gaussian codewords, which is first-order optimal, \cite{polyanskiy2010channel} and \cite{Shannon_1959} employed an input ensemble of codewords from the power shell, 
and Gallager proposed to generate codewords using a truncated Gaussian distribution lying in a thin shell \cite{Gallager_1968}. 
Moreover, the above results can be generalized to the channels with feedback. Specifically, for a fixed error probability and finite average decoding time, Polyanskiy et al. \cite{polyanskiy2011feedback} derived non-asymptotic bounds on the maximum achievable codebook size, 
which were further improved in \cite{yavas2025variable}. 
For discrete memoryless channels and additive white Gaussian noise channels with an average power constraint, the authors in \cite{yavas2025variable} also developed a universal variable-length feedback code that does not rely on the knowledge of the underlying channel parameters. 

2)~Converse Bounds: 
Many converse bounds have been established in the literature. 
Fano's inequality is a classic tool to prove the weak converse bound (i.e., upper bound) on \(R^{\star}(n,\epsilon)\) \cite[Sec. 7.9]{cover2006elements}, while Wolfowitz \cite{Wolfowitz} established a strong converse bound for discrete memoryless channels. 
The authors in \cite{Shannon3} derived a sphere-packing converse bound suitable for the analysis of reliability-function. 
Verd\'u and Han \cite{Verdu} established an information-spectrum-based converse bound, which holds for arbitrary random transformations. 
Later, based on hypothesis testing and various data-processing inequalities, Polyanskiy, Poor, and Verd\'u derived a general converse bound, known as the meta-converse bound \cite{Yury_phd}. 
It was shown in \cite[Sec. 2.7.3]{Yury_phd} and \cite{Yury_conf1} that most converse results can be recovered from this meta-converse bound.

\begin{figure}[t]
\centering
\includegraphics[width=0.8\linewidth]{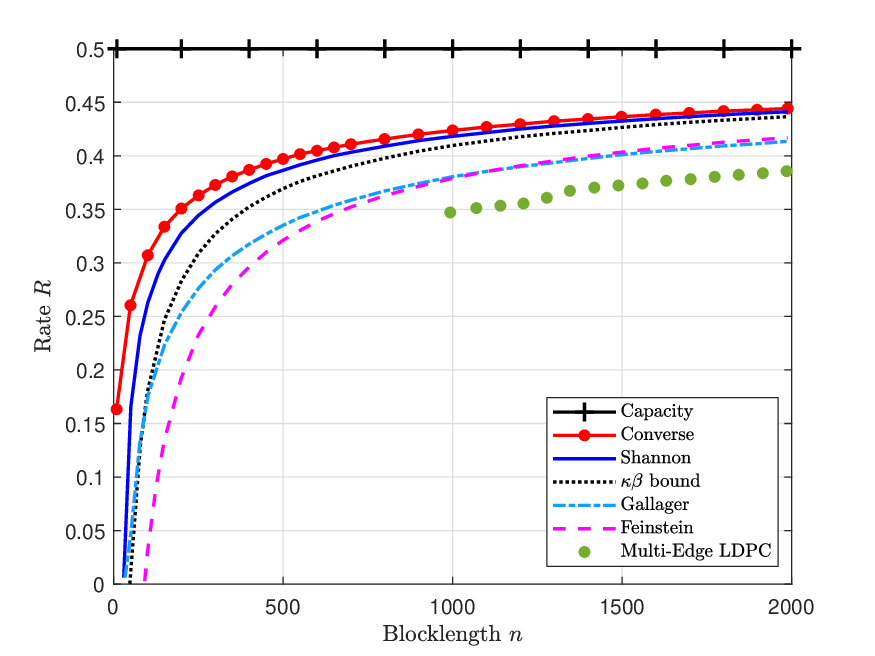}
\caption{Theoretical Bounds and the multi-edge LDPC algorithm for P2P in AWGN channels with SNR \(= 0\)~dB and \(\epsilon = 10^{-3}\).} \label{Fig:awgn}
\end{figure}

3)~Numerical Results: 
Fig.~\ref{Fig:awgn} compares various achievability and converse bounds on \(R^{\star}(n,\epsilon)\) under the maximal power constraint in P2P AWGN channels. 
The large gap between the non-asymptotic bounds and asymptotic capacity reveals the significance of applying finite-blocklength information theory for short-packet communications with small \(n\) and \(\log M\). 
Among non-asymptotic achievability bounds, Shannon's bound demonstrates superior tightness, but is restricted to AWGN-specific configurations. 
Compared with Shannon's bound, the \(\kappa\beta\) bound is slightly looser, but is more computationally tractable for asymptotic analysis and more general. 
The Feinstein bound is looser than the \(\kappa\beta\) bound. 
Compared with Gallager's bound, the \(\kappa\beta\) bound is tighter in large \(n\) regimes, but looser for small \(n\). 
Fig.~\ref{Fig:awgn} also shows the performance of the multi-edge low-density parity-check (LDPC) code with a low-complexity belief-propagation based decoder. 
The comparison between this practical scheme and capacity shows that this scheme becomes closer to optimal as \(n\) increases. 
However, we can observe that the gap between this scheme and finite-blocklength bounds is largely blocklength independent. 
This discrepancy highlights the paradigm shift required in code design evaluation -- moving beyond asymptotic metrics to embrace finite-blocklength information theory for accurate performance assessment in short-packet communications.

The next generation of channel coding is not only required to satisfy the stringent requirements of 6G, but also expected to be backward compatible to avoid imposing additional burden on the crowded baseband chip. Motivated by this, the authors in \cite{ShenLi_Mag} reviewed the potential channel codes for 6G communications, and explored next-generation channel codes based on LDPC and polar frameworks.
A novel concept called generalized LDPC with polar-like component (GLDPC-PC) codes was introduced in \cite{ShenLi_Mag}, where the soft information passed by polar components to variable nodes is efficiently extracted from the soft-output successive cancellation list (SO-SCL) decoder~\cite{YuanPH}. 
Considering that the sequential nature of successive cancellation list decoding leads to a high decoding latency for the SO-SCL decoder, the authors in \cite{ShenLi_WCNC} proposed a soft-output fast SCL (SO-FSCL) decoder by incorporating node-based fast decoding into the SO-SCL framework. 
Interested readers can refer to \cite{ShenLi_Mag,YuanPH,ShenLi_WCNC} for more details.

\subsubsection{Approximations}



It is of great significance to develop tractable approximations on \(R^{\star}(n,\epsilon)\), with lower computational complexity than upper and lower bounds, for providing engineering insights on wireless communication system design. 
This pursuit includes three complementary analytical techniques: the central limit theorem (CLT), large deviations (LD), and moderate deviations (MD), as will be introduced later.

Building on the classical expansion results of Strassen \cite{strassen1962asymptotische}, it was shown by Polyanskiy, Poor, and Verd\'u \cite{polyanskiy2010channel} that in the CLT regime, the maximum coding rate \(R^{\star}(n,\epsilon)\) can be tightly approximated by introducing a second-order statistic of the channel, i.e., the channel dispersion, defined as \cite[Def. 1]{polyanskiy2010channel} 
\begin{equation}
    V = \lim_{\epsilon\to0}  \lim_{n\to\infty} n \left( \frac{C-R^{\star}(n,\epsilon)}{Q^{-1}(\epsilon)} \right)^2 .
\end{equation}
For various channels with a positive capacity \(C\), the maximum coding rate \(R^{\star}(n,\epsilon) \) is tightly approximated by normal approximation~\cite{polyanskiy2010channel}
\begin{align}
    R^{\star}(n,\epsilon) 
    & = C - \sqrt{\frac{V}{n}} Q^{-1}(\epsilon) + O\left( \frac{\log n}{n} \right) . 
\end{align}
This approximation indicates that to guarantee the error probability below \(\epsilon\) with blocklength \(n\), one pays a penalty on the rate (compared to the channel capacity) that is proportional to \(\frac{1}{\sqrt{n}}\). 
As the blocklength \(n\) goes to infinity, the rate penalty tends to \(0\).

In AWGN channels with signal-to-noise ratio (SNR) \(P\), \(C\) and \(V\) are given by~\cite[Theorem 54]{polyanskiy2010channel}
\begin{equation}
    C = \frac{1}{2}\log (1+P) , 
\end{equation}
\begin{equation}
    V = \frac{P}{2} \frac{(2+P)}{(1+P)^2} \log^2 e. 
\end{equation}
The results in \cite{polyanskiy2010channel} have been generalized to various wireless communication channels. 
In MIMO quasi-static fading channels where fading coefficients remain invariant over the duration of a codeword, under mild conditions on the fading distribution as shown in \cite[Theorem 3]{Yang_quasi}, the channel dispersion was proved to be zero, i.e., it is satisfied that \cite{Yang_quasi}
\begin{equation}
    R^{\star}(n,\epsilon) = C_{\epsilon}  + O\left( \frac{\log n}{n} \right). \label{eq:Yang_MIMO}
\end{equation}
The channel dispersion has also been studied for ergodic fading channels. 
Specifically, the dispersion of single-input single-output (SISO) stationary fading channels with known channel state information (CSI) at the receiver was derived in \cite{Yury_ISIT}, which was extended to SISO block-memoryless fading channels in 
\cite{Yang_ITW} and for multiple-input multiple-output~(MIMO) block-memoryless fading channels in \cite{Collins}. 
For the asymptotically ergodic setup where the number of antennas grows linearly with the blocklength, upper and lower bounds on the second-order coding rate in MIMO quasi-static Rayleigh fading channels were provided in \cite{Hoydis}. 
You et al. \cite{You_TWC} derived a closed-form approximation to the upper bound on the achievable rate in massive MIMO systems as follows
\begin{equation}
    \frac{R^{\star}(n,\epsilon)}{m} \leq \log(1+P) - \sqrt{\frac{1}{nm}} Q^{-1}(\epsilon), \label{eq:You_massiveMIMO}
\end{equation}
where $m$ denotes the spatial degree of freedom. 
It is shown that the average data rate per Hz per antenna remains unchanged when we fix the value of $mn$, revealing that $n$ and $m$ can be thought of being reciprocal to some extent and thus the required blocklength can be greatly reduced by increasing the number of antennas.
Subsequently, the explicit performance bounds for spatiotemporal coding were derived in \cite{You_TIT}, which emerges as a critical enabler for latency reduction in 6G systems \cite{You_SCIS}.

The analysis of the tradeoff between error probability, rate, and blocklength in the LD regime dates back to Gallager's pioneering work in 1960~\cite{Gallager_1965}. 
It was proved that for a fixed rate \(R\) strictly below capacity, the error probability \(\epsilon^{\star}(n,R)\) decays exponentially with \(n\), i.e., 
\begin{equation}
    \epsilon^{\star}(n,R) = \exp(-n ( E(R) + o(1) ) ). 
\end{equation}


The CLT- and LD-type approximations are dominant in different cases: CLT approximations dominate if \(\epsilon\) is large (i.e. \(R\) is close to the capacity); and LD approximations dominate if \(\epsilon\) is small (i.e. \(R\) is much smaller than the capacity).  
The MD analysis is performed between the above two regimes \cite{Yury_conf2, Altug}, which shows that the error probability decays subexponentially with \(n\) and the maximal achievable rate converges to the channel capacity slower than \(1/\sqrt{n}\). Specifically, the MD analysis yields the following approximation
\begin{equation}
    \epsilon^{\star}(n,C-\delta_n) \approx Q\left( \sqrt{\frac{n}{V}} \delta_n \right) \approx \exp \left( - \frac{n \delta_n^2}{2V} \right), 
\end{equation}
where \(\delta_n>0, \delta_n\to0\), and \(n \delta_n^2\to\infty\) as \(n \to \infty\). 
A third-order approximation on the maximum rate was derived in~\cite{yavas2024third} based on the MD analysis. 

\subsection{Multiple Access Channels}
\label{subsec:mac cc}
The multiple access channel (MAC), in which multiple users access the system simultaneously, is a foundational model in wireless communication networks. 
For a Gaussian MAC with a fixed number \(K\) of users, the asymptotic capacity region is defined as the convex hull of all achievable rate tuples \((R_1, R_2, \dots, R_K)\) satisfying 
\begin{equation}
    \sum_{i \in S} R_i \leq C \left( \sum_{i \in S} P_i \right), \quad \forall S \subseteq \{1, 2, \dots, K\}, 
\end{equation} 
where \(P_i\) denotes the SNR of user \(i\) and \(C(\cdot)\) denotes the Shannon capacity of the P2P Gaussian channel. 
This region is a \(K\)-dimensional polytope bounded by \(2^K - 1\) inequalities, reflecting the trade-off between individual rates and their sum. 
In the case of \(K=2\), the capacity region simplifies to a pentagon~\cite{Wyner}. 


The above capacity analysis relies on the assumption of infinite blocklength. For MAC with finite blocklength satisfying \(K \ll n\), MolavianJazi and Laneman \cite{MolavianJazi} and Huang and Moulin \cite{huang2012finite} generalized the finite-blocklength result for the P2P channel to the two-transmitter MAC, where the finite-blocklength results were established based on the global dispersion matrix. 
As discussed by Haim et al. \cite{haim2012note}, in most cases of interest, non-diagonal entries of the dispersion matrix do not play a fundamental role in characterizing the performance, especially when approaching a point on the surface of the rate-region which is not a corner. This motivates the study of local second-order asymptotics for the multiple access channels, as shown in \cite{Scarlett,scarlett2015second}.

%

%

\subsection{Massive Random Access Channels}
\label{subsec:mra cc}
\subsubsection{Information-Theoretic Bounds}

The rapid expansion of Internet of Things (IoT) applications has elevated ultra-massive machine-type communication (umMTC) to a pivotal role in next-generation wireless systems. 
Compared with conventional MACs, where the number of users is usually fixed and much smaller than the blocklength, a key challenge in umMTC is to enable efficient and reliable random access for a large number of users (comparable to or larger than the blocklength), among which only a fraction is active and transmits a short packet (e.g., several hundred bits) to the base-station~(BS) under limited channel uses and stringent energy constraints~\cite{wuyp,liu2018massive,liu2018sparse}.
Two distinct massive random access paradigms have emerged: 1) individual codebook-based massive random access, where each user is assigned a unique codebook for activity detection and message decoding; and 2) common codebook-based massive random access, where users share a common codebook, and the receiver recovers a permutation-invariant list of transmitted messages without associating them with specific users. The second one is also termed unsourced random access (URA)~\cite{A_perspective_on}.
Characterizing the fundamental limits of massive random access is of great significance.

A model called many-access channel~(MnAC) was proposed in~\cite{GuoDN} to characterize the massive user population, in which the number of users grows with the blocklength. 
This model was adopted in subsequent studies under the assumption of linear scaling. 
However, the work~\cite{GuoDN} overlooked practical constraints like finite energy-per-bit, payload size, and blocklength. 
To this end, some works considered the regime with infinite blocklength but finite payload size and energy-per-bit~\cite{improved_bound,finite_payloads_fading,GaoJY}, where the per-user probability of error (PUPE) criterion proposed in~\cite{A_perspective_on} was adopted. 
Particularly, based on the MnAC model with linear scaling, assuming each user is allocated with an individual codebook and the BS is equipped with a single antenna, Zadik et al.~\cite{improved_bound} and Kowshik et al. \cite{finite_payloads_fading} derived achievability and converse bounds on the minimum required energy-per-bit in AWGN channels and quasi-static fading channels, respectively, revealing that multi-user interference~(MUI) can be almost perfectly canceled at low user densities.

Latency-critical applications necessitate finite-blocklength analysis. Non-asymptotic bounds for URA were derived in~\cite{A_perspective_on} and \cite{RAC_fading} for Gaussian and Rayleigh fading channels, respectively. 
The authors in \cite{han2025finite} extended the results in~\cite{A_perspective_on} to the case where each user generates a packet independently and asynchronously according to identical renewal processes.
These works rely on the assumption of knowing the number of active users in advance. 
However, in massive random access channels, users typically have intermittent or bursty communication patterns and access the network without a grant, thereby leading to variations and uncertainty in the number of active users.
To address this, \cite{noKa} considered massive random access in Gaussian channels with a random and unknown number of active users, and derived non-asymptotic bounds on the misdetection and false-alarm probabilities.

The above theoretical results were established for the scenario where the BS is equipped with a single antenna. 
For user activity detection, it was proved in~\cite{Caire1} that with \(n\) channel uses and a sufficiently large antenna array size \(L\) satisfying \(K_a/L = o(1)\), the BS can detect up to \(K_a = O(n^2)\) active users among \(K\) potential users under the condition of \(\frac{K_a}{K} = \Theta(1)\). This represents a substantial improvement over single-antenna systems, where the number of active users is only allowed to scale linearly with the blocklength.  
Motivated by the great potential of multiple receive antennas, the authors in~\cite{myTIT} derived non-asymptotic achievability and converse bounds for massive random access with individual codebooks over MIMO quasi-static Rayleigh fading channels. 
Based on these results, the fundamental trade-offs between blocklength, payload size, user density, the number of receive antennas, and error probability were characterized, in both cases with and without known CSI at the receiver. 
It was proved in~\cite{myTIT} that in the case with unknown CSI, under the PUPE criterion, when the number of receive antennas is \(L=\Theta\left(n^2\right)\) and the transmit power is \(P=\Theta\left(\frac{1}{n^2}\right)\), one can reliably serve up to \(K=O(n^2)\) users. 
Under mild conditions in the case with known CSI, the PUPE requirement is satisfied if and only if \(\frac{nL\ln{KP}}{K}=\Omega(1)\). This condition highlights the potential of MIMO technology in enabling low-cost and low-latency communications for massive IoT applications. 
It was demonstrated that the energy-per-bit can be finite and even approach zero under certain conditions, which is crucial for practical systems with limited energy resources.  
Moreover, the fundamental limits of URA in MIMO channels were explored in \cite{letter} and \cite{Gao_TIT2}, in which finite-blocklength bounds for the cases with and without known number of active users were established, respectively.

\subsubsection{Comparison with Practical Schemes}

\begin{figure}[t]
\centering
\includegraphics[width=0.8\linewidth]{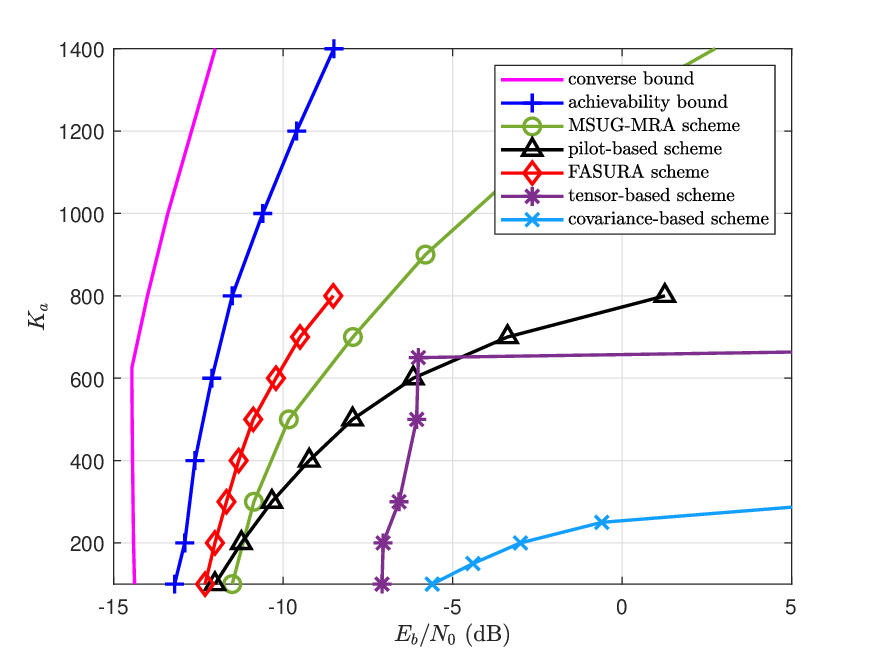}
\caption{Comparison of existing schemes and theoretical achievability and converse bounds for URA in MIMO quasi-static fading channels under the assumption that \(K_a\) is fixed and known in the case of \(n=3200\), \(J = 100\)~bits, \(L=50\), and \(\epsilon_{\rm MD} = \epsilon_{\rm FA} = 0.025\).} \label{Fig:MIMO_scheme}
\end{figure}

The established finite-blocklength bounds serve as theoretical benchmarks for evaluating practical schemes. 
In Fig.~\ref{Fig:MIMO_scheme}, we compare finite-blocklength achievability and converse bounds (using codewords uniformly distributed on a sphere) for URA in MIMO quasi-static fading channels derived in~\cite{letter}, as well as existing state-of-the-art schemes proposed in~\cite{Caire1,Duman2,Caire2,Fasura,TBM}, in the setting with blocklength \(n=3200\), payload size \(J=100\)~bits, the number of BS antennas \(L=50\), and target misdetection and false-alarm probabilities \(\epsilon_{\rm MD} = \epsilon_{\rm FA} = 0.025\).
Key implementation details of state-of-the-art schemes are as follows. 
The MSUG-MRA scheme is evaluated as in~\cite[Fig.~4]{Duman2}, which implements a slotted structure dividing \(n\) into \(S\) slots. Each active user randomly selects a single slot, where the transmitted \(J\)-bit message is divided into \(D\) orthogonal pilot segments (each \(B_p\)~bits) and a \(J-DB_p\)~bit coded segment, enhanced through \(G\) distinct interleaver-power group allocations.
The pilot-based scheme is evaluated as in~\cite[Fig.~7]{Caire2}, which divides data into two parts with \(16\)~bits and \(84\)~bits, respectively. The first part features ``pilot'' of length \(1152\) and the second one is coded by a polar code of length \(2048\). 
The FASURA scheme is evaluated as in~\cite[Fig.~4]{Fasura}, where the message is divided into two parts as in the pilot-based scheme. Departures from~\cite{Caire2} include the use of spreading sequences of length \(L=9\), the method of detecting active sequences, and channel/symbol estimation techniques. 
The tensor-based scheme is evaluated as in~\cite{TBM}, which utilizes the tensor signature \((8,5,5,4,4)\) with BCH outer coding, tolerating error probability \(\epsilon=0.1\). 
The covariance-based scheme was proposed in~\cite{Caire1}. We employs 16-slot frames, each of length 200 including 15 bits with the tree code parity profile \( [0,7,8,8,9, \ldots, 9, 13 , 14] \). 
We can observe from Fig.~\ref{Fig:MIMO_scheme} that existing schemes maintain competitive energy efficiency when \(K_a\) is small, but suffers from more performance degradation and requires higher energy-per-bit \(E_b/N_0\) compared with theoretical bounds as \(K_a\) increases, calling for more advanced and energy-efficient scheme design in the case with large \(K_a\).

\section{Joint Source and Channel Coding}
\label{sec:jscc}

\begin{figure}[t]
  \centering
  \includegraphics[width=0.7\linewidth]{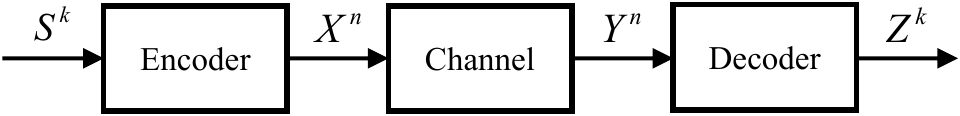}
  \caption{System model for joint source-channel coding.} \label{Fig:jscc}
\end{figure}

For clarity, we present the JSCC system model in Fig.~\ref{Fig:jscc}. The encoder input and decoder output are the \(k\)-length vectors \(S^k\) and \(Z^k\). The \(n\)-length vectors \(X^n\) and \(Y^n\) are the noisy channel input and output, respectively. The JSCC coding rate is defined as \(\frac{k}{n}\). In the asymptotic regime, as the excess-distortion probability vanishes, the maximum achievable JSCC coding rate is \(\frac{C}{R(d)}\), where \(C\) is the \textit{channel capacity} and \(R(d)\) is the \textit{rate-distortion function} under a preset average distortion constraint \(d\). This is the classic result from the source-channel separation theorem in~\cite{shannon1959coding} and~\cite{cover2006elements}. In other words, in the asymptotic regime, one designs optimal source and channel codes separately and then concatenates them to achieve the fundamental asymptotic limit. However, in the non-asymptotic regime, the SSCC yields very weak non-asymptotic bounds. In contrast, the JSCC design provides significant gains at finite blocklengths~\cite{kostina2013lossyJSCC}. This implies that the two designs have different theoretical limits and that further research in the finite blocklength regime is essential.

Csiszár in~\cite{csiszar1982error} proved that the error exponent for JSCC exceeds that for SSCC. For discrete source-channel pairs under an average distortion criterion and for transmitting a Gaussian source over a discrete channel under an average mean square error constraint, Pilc in~\cite{pilc1967coding,pilc1968transmission} and Wyner in~\cite{wyner1968communication,wyner1972transmission} obtained non-asymptotic bounds, respectively. Csiszár also derived non-asymptotic fundamental limits for a graph-theoretic model of JSCC in~\cite{csiszar1981graph}. In an almost-lossless JSCC setting, Campo et al. in~\cite{campo2011random} presented several finite blocklength random-coding bounds. Later, Wang et al.~\cite{wang2011dispersion} determined the dispersion for JSCC with finite source and channel alphabets. Finally, Kostina et al. in~\cite{kostina2013lossyJSCC} derived new general achievability and converse bounds for JSCC, provided a Gaussian approximation analysis, and obtained results for various cases. In the following, we review the main results regarding the fundamental limits of JSCC, as well as the basic ideas behind them. We then discuss the comparison between SSCC and JSCC, highlighting the finite-blocklength performance differences, and conclude with a summary of recent developments in JSCC for practical systems.

1)~Converse Bounds: 
On one hand, a general converse bound can be derived based on the information random variable. Specifically, by considering the difference between the relative information 
\(\imath_{Y|X||\bar{Y}}(x;y)=\log\frac{dP_{Y|X=x}}{dP_{\bar{Y}}}(y)\) 
and the \(\mathsf{d}\)-tilted information \(\jmath_S(S,d)\) in~\eqref{eq:d_tilted_information}, it was shown that the probability 
\(\mathbb{P}\left[\jmath_S(S,d)-\imath_{Y|X||\bar{Y}}(X;Y)\ge\gamma\right]\) 
and the term \(e^{-\gamma}\) differ by no more than the excess-distortion probability \(\epsilon\). In addition, this result can be further extended when different channel input types. However, in some cases the bounds produced by this method become somewhat loose.

On the other hand, Csiszár employed a list decoder that produces a list of \(L\) elements from an \(M\)-symbol set to obtain a converse bound for JSCC in~\cite{csiszar1982error}. Kostina et al. in~\cite{kostina2013lossyJSCC} extended this technique from finite alphabet sources to abstract alphabets and combined it with hypothesis testing to obtain a stronger converse bound.

2)~Achievability Bounds: 
Using the optimal SSCC performance as the nonasymptotic bound for JSCC is a natural idea. A practical SSCC method adopts independent random source codes and random channel codes, which yields computable finite blocklength achievable bounds for JSCC. This approach reveals an important insight into SSCC's suboptimality at finite blocklengths.
%
%
First consider how SSCC reaches the asymptotic limit. In the large-blocklength regime, the optimal source encoder produces outputs that are nearly uniform over a set containing roughly \(\exp(kR(d))\) messages, which cover most source outcomes within the prescribed distortion \(d\). According to the channel coding theorem, there exists a channel code with a maximum-likelihood (ML) decoder that can reliably distinguish up to \(M=\exp(kR(d))<\exp(nC)\) messages. Thus, when the optimal source and channel codes are concatenated, the overall scheme achieves a negligible probability of excess distortion provided that \(d>D(nC/k)\). However, when operating at finite blocklength, the distribution produces by the optimal source encoder is not close to uniform. As a result, a separated scheme employing a ML decoder without accounting for the uneven message probabilities would fail to reach near-optimal nonasymptotic performance.


3)~Approximations:
Before introducing the Gaussian approximation of the JSCC coding rate, certain conditions should be specified. The source and channel are stationary and memoryless, the distortion criterion is separable, and the distortion is bounded. If there is a cost function for the channel, it should also be separable. Under these conditions, the parameters of an optimal \((k,n,d,\epsilon)\) code satisfy
\begin{align} \label{eq:jscc_approx}
  nC(\beta) - kR(d) = \sqrt{nV(\beta) + k\mathcal{V}(d)}\,Q^{-1}(\epsilon) + \theta(n),
\end{align} 
where \(\mathcal{V}(d)\) denotes the source dispersion and \(V(\beta)\) represents the channel dispersion, with \(\beta\) as the channel input cost constraint.
For the remainder term \(\theta(n)\), if the channel input \(X^n\) and output \(Y^n\) are defined on finite alphabets \(\mathcal{X}\) and \(\mathcal{Y}\) with no channel cost constraint, then
\begin{equation}
-c \log n + O(1) \le \theta(n) \le C_0 \log n + \log \log n + O(1), 
\end{equation}
where \(c = |\mathcal{X}|-\frac{1}{2}\) and \(C_0 = \frac{1}{2} + \frac{\operatorname{Var}\!\left(J^{\prime}_{\mathsf{Z}^{\star}}(\mathsf{S},\lambda^\star)\right)}{\mathbb{E}\!\left[\lvert J^{\prime\prime}_{\mathsf{Z}^{\star}}(\mathsf{S},\lambda^\star) \rvert\right]\log e}\) with \(\lambda^\star = -\mathbb{R}_S'(d)\). For further results in the Gaussian channel case, please refer to~\cite[Theorem 10]{kostina2013lossyJSCC}.

For the approximation under SSCC, combining the relevant results of channel coding in~\cite{polyanskiy2010channel} and lossy source coding in~\cite{kostina2012fixed}, it is established that
\begin{align} \label{eq:sscc_approx}
nC(\beta)-kR(d) \le \min_{\eta+\zeta\le\epsilon} \biggl[\sqrt{nV(\beta)}\,Q^{-1}(\eta)+\sqrt{k\mathcal{V}(d)}\,Q^{-1}(\zeta)\biggr] + O(\log n).
\end{align}
If either the channel or the source (or both) exhibits zero dispersion, separate coding can achieve the same overall dispersion as a joint design. In such cases, the \(\mathsf{d}\)-tilted information or the channel information density is nearly deterministic so that an optimal joint source-channel code does not need to account for the full variability in these random quantities.

4)~Comparison between SSCC and JSCC: The comparison between the approximation of JSCC in~\eqref{eq:jscc_approx} and the approximation of SSCC in~\eqref{eq:sscc_approx} offers a more intuitive and straightforward explanation of the finite blocklength performance loss due to the separate design in SSCC.
First, consider the behavior of \(\mathsf{d}\)-tilted information and the channel information density under the central limit theorem as \(k\) and \(n\) become large.
Because the source is stationary and memoryless, the normalized \(\mathsf{d}\)-tilted information \(J = \frac{1}{n}\jmath_{S^k}(S^k,d)\) tends toward a Gaussian distribution with mean \(\frac{k}{n}R(d)\) and variance \(\frac{k}{n}\frac{\mathcal{V}(d)}{n}\). Similarly, the conditional normalized channel information density \(I = \frac{1}{n}\imath^\star_{X^n;Y^n}(x^n;Y^{n\star})\) tends toward a Gaussian with mean \(C(\beta)\) and variance \(\frac{1}{n}V(\beta)\) for all \(x^n\) that are typical under the capacity-achieving distribution. Since an efficient encoder selects such inputs for nearly every source realization, and given the independence between the source and the channel, the difference \(I - J\) is itself nearly Gaussian with mean \(C(\beta) - \frac{k}{n}R(d)\) and variance \(\frac{1}{n}\left(\frac{k}{n}\mathcal{V}(d) + V(\beta)\right)\).
Under JSCC, the source is successfully reconstructed within distortion \(d\) if and only if the channel information density \(I\) exceeds the source \(\mathsf{d}\)-tilted information \(J\), i.e., \(\{ I > J \}\), as indicated by~\eqref{eq:jscc_approx}. By contrast, under SSCC, as indicated by~\eqref{eq:sscc_approx}, the reconstruction is successful with high probability only if the pair \((I, J)\) falls within the intersection of the half-planes \(\{ I > r \}\) and \(\{ J < r \}\), where \(r = \frac{\log M}{n}\) represents the capacity of the noiseless channel between the source and channel code blocks.
Because the event \(\{ I > r \} \cap \{ J < r \}\) is strictly contained within \(\{ I > J \}\), this leads to a performance loss in the separate coding design.

5)~Recent developments in JSCC: Recent research on JSCC has highlighted its potential in practical finite-blocklength communication systems. Task-oriented JSCC strategies aim to transmit only the most relevant information for a given task, which is particularly beneficial in low-latency, bandwidth-constrained settings~\cite{gunduz2024joint}. Deep learning-based JSCC schemes have shown that finite-length source sequences can be directly mapped to channel inputs without separate compression and channel coding, achieving smooth performance degradation across varying channel conditions and avoiding the cliff effect that typically arises in short-blocklength separation-based designs~\cite{bourtsoulatze2019deep}. Furthermore, incorporating channel output feedback in JSCC can significantly enhance end-to-end reconstruction quality and reduce transmission delay, demonstrating concrete advantages of modern JSCC techniques in the finite-blocklength regime~\cite{kurka2020deepjscc}.

\section{Open Problems}
\label{sec:open}
While a number of contributions have been made towards finite-blocklength information theory, this topic remains to be further explored within a broader range of scenarios and requirements. In the following, we will discuss some of these open problems and future research directions in details.

\subsection{Tight and Analytically Tractable Non-Asymptotic Results}

The dual pursuit of high precision and analytical tractability remains a central objective in non-asymptotic information theory. While existing studies have derived numerous finite-blocklength bounds and approximations, critical gaps persist in simultaneously achieving rigorous accuracy and analytical tractability in some cases. 

It was recently demonstrated in~\cite{yavas2024third} that for channel coding, a third-order approximation under the moderate deviations regime achieves remarkable accuracy even for ultra-short blocklengths (e.g. \(n=100\)) and ultra-low error probabilities (e.g. \(10^{-10}\)). However, analogous higher-order approximations for rate-distortion problems are still unexplored, which is a promising direction for future research.

Moreover, non-asymptotic achievability and converse bounds on the minimum energy-per-bit required for massive random access were derived in \cite{RAC_fading} for single-receive-antenna fading channels and in \cite{myTIT,letter} for MIMO fading channels. 
However, these results include intricate matrix operations and numerical optimization, resulting in significant computational burdens. This calls for non-asymptotic bounds with lower complexity while preserving analytical tightness. 
Also, it is interesting to introduce some artificial intelligence (AI) technologies~\cite{wang2024review,yue2024federated} into computational processes to enhance efficiency.

\subsection{Non-Asymptotic Results for More General Scenarios}

For the scenario with a single GMS, while the second-order approximations for the rate-distortion and successive refinement problems have been established, the fundamental limits of multiterminal lossy source coding remain open. Specifically, it is highly non-trivial to derive a non-asymptotic achievability bound for the Kaspi problem with a GMS. In the multiple descriptions problem, while a rate-distortion region for a GMS has been established~\cite{ozarow1980source}, deriving non-asymptotic bounds and second-order approximations requires innovative techniques.

Existing non-asymptotic results on channel coding are expected to be extended to more general scenarios. 
For instance, cell-free massive MIMO has been proposed as an advanced technique to support a large number of users in an expanded coverage area. However, the finite-blocklength fundamental limits of channel coding in such distributed antenna systems have not been explored. 
Moreover, most existing results focus on Gaussian channels and i.i.d. fading channels. 
The fundamental limits in correlated channels remain open.

It is of great significance to find the finite-blocklength fundamental limits of joint source and channel coding for a wider class of sources and channels, such as multiple sources, fading channels, and multiple receive antennas, which is an interesting topic in the future. 
Also, the incorporation of semantic information is a promising future research direction~\cite{{ping2024semantic,huanling2023representation}}.



\subsection{Efficient Practical Scheme Design}

The non-asymptotic bounds and approximations serve as theoretical benchmarks for assessing the performance of practical communication schemes. Existing schemes are shown to exhibit substantial gaps to the theoretical results in some cases. 
Thus, it is essential to develop practical schemes that are closer to the theoretical bounds while maintaining computational feasibility, which is an interesting topic in the future. 
Successfully bridging this gap is of great significance for supporting latency-critical applications in emerging 6G use cases.

\section{Conclusion}
\label{sec:conclusion}
Classical asymptotic information theory, which relies some ideal assumptions, such as infinite blocklength and payload size and vanishing error probability, has some limitations in characterizing the fundamental limits of practical latency-critical communication systems. 
This has motivated us to explore rigorous non-asymptotic frameworks -- a pursuit demanding novel analytical tools and techniques to address the challenges in the finite-blocklength regime. 
In this paper, we systematically reviewed recent advances in the non-asymptotic fundamental limits. 
Specifically, we presented various non-asymptotic achievability bounds, converse bounds, and approximations to the information-theoretic non-asymptotic fundamental limits, as well as key ideas behind these results. 
We started with the foundational results for source coding, rigorously analyzing both lossless and lossy compression in P2P and multiterminal cases. This exploration encompassed memoryless and memory source models, while addressing scenarios with both perfectly known and statistically mismatched source distributions. 
For channel coding, we discussed finite-blocklength results on the tradeoff between data rate, error probability, and blocklength in P2P systems, followed by recent advances in multiple access channels and emerging massive access channels.  
We further presented non-asymptotic results in joint source and channel coding, which was shown to bring considerable performance advantage over a separate one at finite blocklengths -- a departure from the conclusion in classical asymptotic information theory. 
The paradigm shift moving from asymptotic metrics to finite-blocklength information theory facilitates accurate performance characterization of practically relevant scenarios. 
Also, knowledge of the behavior of the fundamental limits in the non-asymptotic regime enables the assessment of practical schemes, which were shown to exhibit a large gap to the theoretical results in some cases. 
Finally, some open challenges in finite-blocklength information theory were discussed, which are essential for advancing information-theoretic analysis for future wireless communication systems.

\bibliographystyle{elsarticle-num} 
\bibliography{main_references}
\end{document}